\documentclass{nature}
\usepackage[utf8]{inputenc}
\usepackage{graphicx}
\graphicspath{ {./figures/} }
\usepackage{textcomp,gensymb}
\usepackage[euler]{textgreek}
\usepackage{lineno}

\usepackage{caption}
\DeclareCaptionLabelFormat{vertline}{#1 #2 $|$} 
\usepackage{prettyref}
\newrefformat{fig}{Fig.~\ref{#1}}
\newrefformat{exfig}{Supplementary Fig.~\ref{#1}}

\usepackage[normalem]{ulem}

\usepackage{color}

\usepackage{units}
\usepackage{amsmath,amssymb}
\usepackage{comment}

\makeatletter
\let\saved@includegraphics\includegraphics
\AtBeginDocument{\let\includegraphics\saved@includegraphics}
\renewenvironment{figure}{\@float{figure}}{\end@float}

\makeatother

\newcommand{\beginsupplement}{%
        \setcounter{table}{0}
        \renewcommand{\table}{\arabic{table}|}%
        \setcounter{figure}{0}
        \renewcommand{\figurename}{Supplementary Figure}
        \renewcommand{\thefigure}{S\arabic{figure}}%
     }

\begin{document}

\title{Interaction-driven Band Flattening and Correlated Phases in Twisted Bilayer Graphene}

\author{Youngjoon Choi$^{1,2,3*}$, Hyunjin Kim$^{1,2,3*}$, Cyprian Lewandowski$^{2,3,4}$, Yang Peng$^{5}$, Alex Thomson$^{2,3,4}$,  Robert Polski$^{1,2}$, Yiran Zhang$^{1,2,3}$, Kenji Watanabe$^6$, Takashi Taniguchi$^6$, Jason Alicea$^{2,3,4}$, Stevan Nadj-Perge$^{1,2\dagger}$}

\maketitle

\begin{affiliations}
    \item T. J. Watson Laboratory of Applied Physics, California Institute of Technology, 
         1200 East California Boulevard, Pasadena, California 91125, USA
    \item Institute for Quantum Information and Matter, California Institute of Technology, Pasadena, California 91125, USA
    \item Department of Physics, California Institute of Technology, Pasadena, California 91125, USA
    \item Walter Burke Institute for Theoretical Physics, California Institute of Technology, Pasadena, California 91125, USA
    \item Department of Physics and Astronomy, California State University, Northridge, California 91330, USA
    \item National Institute for Materials Science, Namiki 1-1, Tsukuba, Ibaraki 305 0044, Japan
    \item[*] These authors contributed equally to this work
    \item[$^\dagger$] Correspondence: s.nadj-perge@caltech.edu 
\end{affiliations}

\begin{abstract}

Flat electronic bands, characteristic of magic-angle twisted bilayer graphene (TBG), host a 
wealth of correlated phenomena. 
Early theoretical considerations\cite{bistritzerMoireBandsTwisted2011,lopesdossantosGrapheneBilayerTwist2007} 
suggested that, at the magic angle, the Dirac velocity vanishes and the entire width of the moir\'e bands 
becomes extremely narrow. Yet, this scenario contradicts experimental 
studies\cite{caoCorrelatedInsulatorBehaviour2018, 
kerelskyMaximizedElectronInteractions2019,choiElectronicCorrelationsTwisted2019, 
xieSpectroscopicSignaturesManybody2019, jiangChargeOrderBroken2019, 
tomarkenElectronicCompressibilityMagicAngle2019, nuckollsStronglyCorrelatedChern2020, 
choiCorrelationdrivenTopologicalPhases2021, parkFlavourHundCoupling2020} that reveal a finite Dirac velocity
as well as bandwidths significantly larger than predicted. 
Although more 
sophisticated modeling\cite{uchidaAtomicCorrugationElectron2014, 
jungInitioTheoryMoir2014, namLatticeRelaxationEnergy2017, 
carrExactContinuumModel2019, guineaContinuumModelsTwisted2019} can, in part, 
account for the bandwidth 
broadening, many essential aspects of magic-angle TBG bands and emerging 
correlated phenomena remain elusive. 
Here we use spatially resolved spectroscopy
in finite and zero magnetic fields 
to examine the electronic structure of moir\'e bands and their intricate connection to correlated phases. 
By following the relative shifts of Landau levels in finite fields, we detect filling-dependent band 
flattening caused by strong interactions between electrons, that unexpectedly starts already 
at $\mathbf{\sim1.3\degree}$, well above the magic angle and hence nominally in the weakly correlated regime. 
We further show that, as the twist angle 
is reduced, the moir\'e bands become maximally flat at progressively lower doping levels. 
Surprisingly, when the twist angles reach values for which the maximal flattening occurs at approximate 
filling of $\mathbf{-2}$, $\mathbf{+1}$,$\mathbf{+2}$,$\mathbf{+3}$ electrons per moir\'e unit cell, 
the corresponding zero-field correlated phases start to emerge.  
Our observations are corroborated by calculations that incorporate an interplay between the 
Coulomb charging energy and exchange interactions; together these effects produce band 
flattening and hence a significant density-of-states enhancement that facilitates the 
observed symmetry-breaking cascade transitions. Besides emerging phases pinned to integer fillings, 
we also experimentally identify a series of pronounced correlation-driven band 
deformations and soft gaps in a wider doping range around $\mathbf{\pm2}$ filling where 
superconductivity is expected. Our results highlight the essential role of 
interaction-driven band-flattening in defining electronic properties 
and forming 
robust correlated phases in TBG.
\end{abstract}

\begin{figure}[ht]
\captionsetup{format=plain,labelsep=space}
\begin{center}
    \includegraphics[width=14cm]{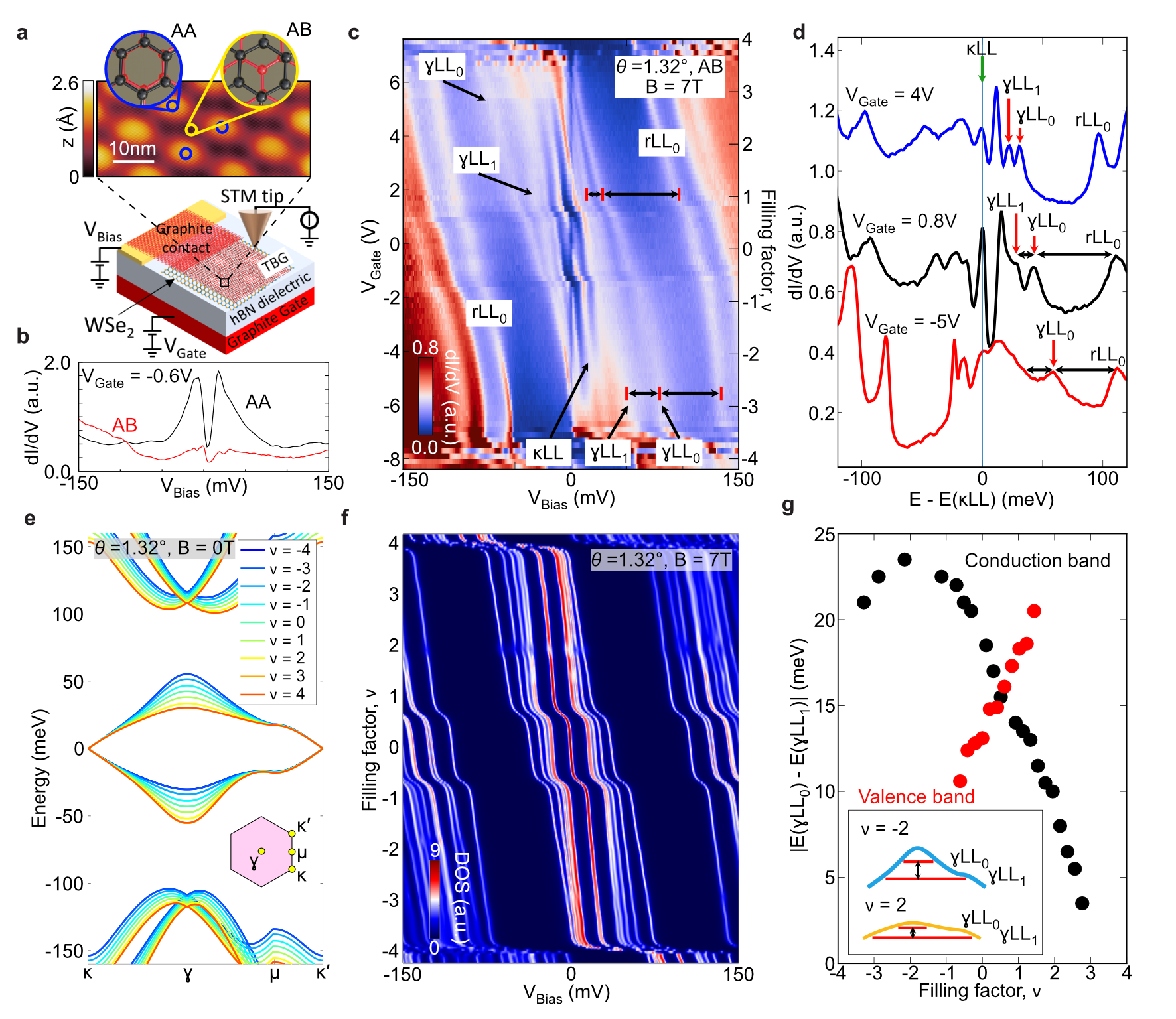}
\end{center}
\caption{{\bf Filling-dependent band structure deformation of TBG at twist 
angle $\mathbf{\theta=1.32\degree}$.} 
{\bf a}, Device schematics and a TBG surface topography. TBG is placed on a monolayer 
WSe$_2$, thin hBN layer and graphite back gate.  A bias voltage $\mathrm{V_{Bias}}$ is 
applied through a graphite contact placed on top. Blue and yellow circles respectively indicate AA- and AB/BA-stacked regions in the 
TBG moir\'e pattern (tunneling set point parameters: 
$\mathrm{V_{Bias} = 100}$~mV, $\mathrm{I = 20}$~pA). {\bf b}, Point spectroscopy at $\mathrm{B = 0}$~T
near the CNP taken at an AA and an AB site; 
AA sites show large LDOS peaks corresponding to VHSs. {\bf c}, Tunneling conductance ($\mathrm{dI/dV}$) 
spectroscopy on an AB site as a function of $\mathrm{V_{Gate}}$ at a magnetic field of $\mathrm{B = 7}$~T ($\mathrm{T = 2}$~K) showing the evolution of LLs
with electrostatic doping. The LLs originating from $\gamma$  and $\kappa$ pockets 
($\gamma$LLs and $\kappa$LLs) of the flat bands as well LLs from remote bands ($r$LLs) are 
identified. %
The energy separation between different 
LLs, as marked by black lines, changes with $\mathrm{V_{Gate}}$. See SI, section 2, 
for conversion between $\mathrm{V_{Gate}}$ and $\nu$. {\bf d}, Linecuts of data in ({\bf c}) at $\mathrm{V_{Gate} = 4}$~V, 1~V, -5~V further illustrate the 
LL spectrum and its change with electrostatic doping.  {\bf e}, 
Calculated TBG band structure with Hartree corrections for $\theta=1.32\degree$ and $\mathrm{B=0}$~T. 
Electron doping flattens the conduction band while hole doping flattens the valence band. 
{\bf f}, Calculated density of states with Hartree corrections as a function of filling for $\mathrm{B=7}$~T (see SI, sections 4 and 5). {\bf g}, Measured energy separation 
between $\gamma$LL$_0$ and $\gamma$LL$_1$ as a function of filling factor showing conductance 
(valence) band flattening for electron (hole) doping. 
}
\label{fig: fig1}
\end{figure}

Figure~\ref{fig: fig1}a sketches our scanning tunneling microscopy (STM) setup. TBG is placed 
on an atomically smooth dielectric consisting of monolayer WSe$_2$ and a thicker ($\sim 30$~nm) layer 
hexagonal boron nitride (hBN)\cite{choiCorrelationdrivenTopologicalPhases2021} (see also Supplementary Information (SI), section 1). By applying a gate 
voltage $\mathrm{V_{Gate}}$ on a graphite gate underneath, we tune the TBG charge density, or 
equivalently filling factor $\nu$ corresponding to the number of electrons per moir\'e 
unit cell. Typical TBG topography shows a moir\'e superlattice consisting of AA sites, where 
the local density of states (LDOS) originating from bands closest to the Fermi energy at charge 
neutrality is predominantly concentrated\cite{kerelskyMaximizedElectronInteractions2019,choiElectronicCorrelationsTwisted2019, 
xieSpectroscopicSignaturesManybody2019, jiangChargeOrderBroken2019}, and AB sites in between 
(\prettyref{fig: fig1}a). Local twist angle and strain are determined by measuring distances 
between neighboring AA sites\cite{kerelskyMaximizedElectronInteractions2019, choiElectronicCorrelationsTwisted2019} (see also SI, section 2). 
We first focus on a TBG region with local twist angle $1.32\degree$ to show that interactions play an important role even well above the magic angle of $\sim 1.1\degree$. %

To further examine the moir\'e band structure, we probe Landau levels (LLs) that 
develop when an out-of-plane magnetic field is applied. The tunneling conductance 
spectrum taken on an AB site shows two different sets of LLs observed as LDOS peaks 
separated by 
the VHSs (\prettyref{fig: fig1}c, d). The LLs from the inner set, with 
energies bounded within the two VHSs\cite{choiCorrelationdrivenTopologicalPhases2021, 
nuckollsStronglyCorrelatedChern2020}, originate from band pockets around 
the $\kappa$ and $\kappa'$ high symmetry points of the moir\' e Brillouin zone; we 
therefore denote them as $\kappa$LLs (see \prettyref{fig: fig1}e). 
Similarly we define LLs from the outer set as $\gamma$LLs since they originate from 
portions of the bands around the $\gamma$ point. This assignment is justified by the 
magnetic-field dependence of the observed LL spectrum (see also \prettyref{exfig: gamma_ll_spectrum}). 
In particular, upon increasing the magnetic field 
the zeroth $\gamma$LLs ($\gamma$LL$_0$ in the valance and 
conduction bands)  
approach the VHSs, 
as expected from the conduction- and valence-band dispersion at the $\gamma$ point;
the $\kappa$LL energies, in contrast, do not change---consistent with the zeroth LLs expected from the 
Dirac-like dispersion at the $\kappa,\kappa'$ points. Moreover, even though both $\kappa$LLs and $\gamma$LLs 
are visible on the AB sites, only the $\kappa$LLs are resolved on AA sites (\prettyref{exfig: gamma_spatial}a, b).
This observation suggests that the spectral weight of the $\kappa,\kappa '$ pockets 
is spatially 
located predominantly on AA sites while the weight of the $\gamma$ pocket is more distributed over AB sites and domain 
walls, in line with previous theoretical calculations\cite{rademakerChargetransferInsulationTwisted2018a, 
guineaElectrostaticEffectsBand2018, carrDerivationWannierOrbitals2019}.

Importantly, the energy separation between $\gamma$LL$_0$ and $\gamma$LL$_1$ changes 
with carrier density (\prettyref{fig: fig1}c, d), signaling significant deformation of 
the moir\'e bands upon doping even at the large 1.32$\degree$ twist angle. For the 
conduction band, the separation is maximized at $\nu=-4$, where 
$E(\gamma$LL$_0)-E(\gamma$LL$_1) \approx 25$~meV, 
and monotonically decreases below $5$~meV near $\nu=+4$ (\prettyref{fig: fig1}g). 
For the valence band, the trend is reversed: the separation between 
$\gamma$LLs increases with filling factor (see \prettyref{exfig: displacement_field}a, b, on remote bands and extracted effective mass). Note that a displacement field, likely 
present due to the single back-gate geometry of our device, might also slightly 
modify the band structure with doping. However, the displacement field would change 
the conduction and the valence bands symmetrically, in contrast to the observed 
asymmetric evolution of $\gamma$LLs (\prettyref{exfig: displacement_field}c, d and SI, sections 3, 4 and 5).

The relative $\gamma$LL shifts upon doping are well-captured within a 
model that includes interaction effects deriving from the inhomogeneous 
charge distribution in a moir\'e unit 
cell\cite{guineaElectrostaticEffectsBand2018,rademakerChargeSmootheningBand2019,
ceaElectronicBandStructure2019,goodwinHartreeTheoryCalculations2020,
calderonInteractions8orbitalModel2020}. 
Starting from charge neutrality, electrons are first added or removed 
from states near $\kappa,\kappa'$  that localize primarily on AA 
sites---creating an associated inhomogeneous electrostatic Hartree 
potential peaked in the AA regions. States near the $\gamma$ point 
feel this potential less strongly because they are relatively 
delocalized within the unit cell, which in turn renormalizes the 
energy cost for populating states near $\gamma$. More generally, each 
part of the moir\'e bands experiences a different electrostatic 
potential, creating  filling-dependent band deformations. 
Figure~\ref{fig: fig1}e presents the calculated band structures for 
different integer fillings at $\mathrm{B = 0}$~T 
(see SI, section 3 for details of the calculation). 
The conduction 
and valence bands deform asymmetrically upon doping: the conduction 
(valence) band becomes flatter and the valence (conduction) band more 
dispersive as the filling factor increases (decreases). Consequently, 
in finite magnetic fields the energy separation between $\gamma$LLs 
also changes asymmetrically (\prettyref{fig: fig1}g inset). 
The Landau level spectrum evaluated with only the electrostatic Hartree 
potential (\prettyref{fig: fig1}f; see also SI, section 5) 
indeed reproduces the main features 
of the experimental data.

\begin{figure}[ht]
\begin{center}
    \includegraphics[width=13.7cm]{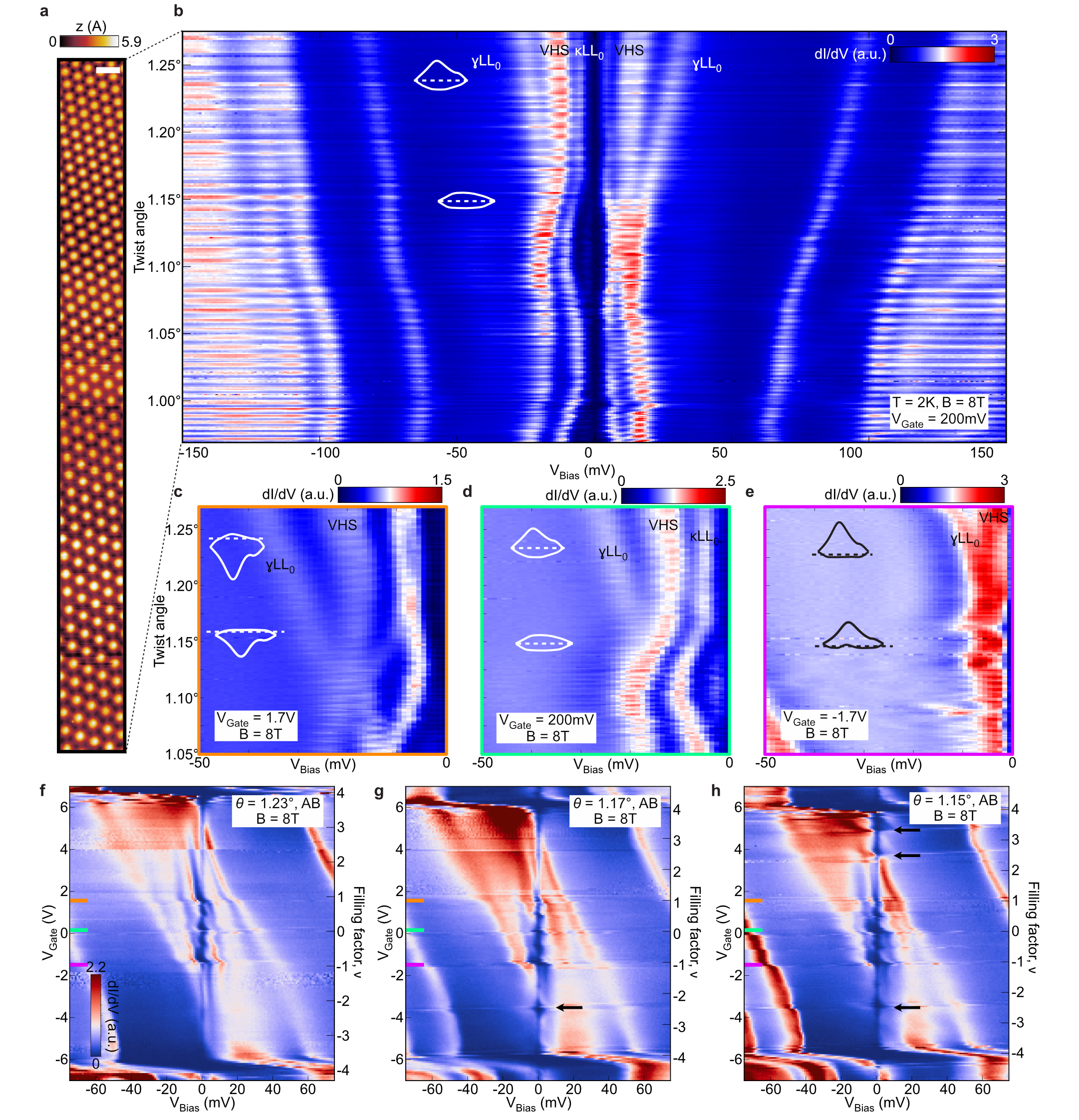}
\end{center}
\caption{{\bf Evolution of LLs with twist angle and correlated gaps at $\mathbf{B = 8}$~T.} 
{\bf a}, A 590 nm $\times$ 50 nm area where twist angle slowly changes from 
$1.27\degree$ to $0.97\degree$ (scale bar corresponds to 20 nm, set point parameters: 
$\mathrm{V_{Bias} = 100}$~mV, $\mathrm{I = 20}$~pA). 
{\bf b}, spectroscopic map near CNP ($\mathrm{V_{Gate}=0.2}$~V) 
taken over the same area, averaged along the horizontal axis while the vertical axis is converted 
into the local twist angle. Evolution of LLs from the flat ($\mathrm{\lvert V_{Bias} \rvert < 50}$~mV) and remote bands ($\mathrm{\lvert V_{Bias} \rvert > 50}$~mV) is clearly resolved.
{\bf c-e}, The same plot as ({\bf b}) focusing on the evolution of the valence-band $\gamma$LL$_0$ for: ($\mathrm{V_{Gate}=1.7}$~V ({\bf c}), electron doping; $\mathrm{V_{Gate}=0.2}$~V ({\bf d}), near the CNP; $\mathrm{V_{Gate}=-1.7}$~V ({\bf e}), hole doping). 
Merging between $\gamma$LL$_0$ and VHS occurs at higher twist 
angle as $\mathrm{V_{Gate}}$ is reduced.
The insets in ({\bf b-e}) sketch 
the band structure and Fermi level near $\theta=1.23\degree$ and $\theta=1.15\degree$. 
A smooth signal background is subtracted to 
enhance LL visibility. 
{\bf f-h}, Point spectroscopy for $\theta = 1.23\degree$ ({\bf f}), $1.17\degree$ ({\bf g}), and 
$1.15\degree$ ({\bf h}). 
Black arrows in ({\bf g}) and ({\bf h}) indicate emerging correlated Chern phases\cite{nuckollsStronglyCorrelatedChern2020,choiCorrelationdrivenTopologicalPhases2021} after the $\gamma$LL$_0$ merges with the VHS. Color coded lines show $\mathrm{V_{Gate}}$ values used in ({\bf c-e}). }
\label{fig: fig2}
\end{figure}

Interaction-induced deformation of the moir\' e bands completely flattens 
the $\gamma$ pocket when the twist angle approaches the magic-angle value, 
as can be deduced from the measured evolution 
of $\gamma$LLs (\prettyref{fig: fig2}). To explore the twist-angle dependence, 
we focus on an area where the angle changes over a
$\sim 600$~nm area and strain is relatively low ($<0.3\%$); see \prettyref{fig: fig2}a. 
Spatially resolved measurements in this area reveal the twist angle-dependent 
evolution of LLs with electrostatic doping (\prettyref{fig: fig2}b-e). At $\mathrm{V_{Gate}}$ near the charge neutrality point (CNP) (\prettyref{fig: fig2}b), we observe that $\gamma$LL$_0$, for both the valence and conduction moir\'e 
bands, eventually merges with the corresponding VHS---signaling maximal band flattening. The onset of merging 
occurs at somewhat larger angles for the valence band ($\sim1.15\degree$) compared to the conduction band 
($\sim1.1\degree$). Moreover, doping strongly shifts this onset. For example, in the valence band 
(\prettyref{fig: fig2}c-e) the maximal band flattening moves considerably towards larger (smaller) twist angles for hole 
(electron) doping (see \prettyref{exfig: gamma_mapping_conduction} for the conduction band).

More detailed examples of the evolution of LLs, along with the development of correlated gaps in finite fields, can be seen in the doping vs.~bias 
maps of \prettyref{fig: fig2}f-h (See also \prettyref{exfig: chern_example} for more data). At $1.23\degree$ (\prettyref{fig: fig2}f), the $\gamma$LL$_0$ energies are well-resolved in both the valence and 
conduction bands between $\nu\approx-2.2$ and 
$\nu\approx+3.3$, beyond which one of the two merges with the corresponding VHS (and also Fermi energy $V_{\rm{Bias}}=0$~mV; see discussion 
of \prettyref{fig: fig3} below). The difference in filling factors of the merging points for the conduction and valence bands 
reflects an appreciable electron-hole asymmetry. 
At this large twist angle, aside from quantum Hall ferromagnetism in the zeroth $\kappa$LL (responsible for the structure between $\nu = -1$ and $+1$), no pronounced correlated gaps are observed at the Fermi energy. %
As the twist angle is decreased to $1.17\degree$, $\gamma$LL$_0$ in the valence band merges at a lower $\lvert \nu \rvert$, and an additional
correlated gap appears when the merged $\gamma$LL$_0$ crosses the Fermi energy (black arrow in \prettyref{fig: fig2}g). For an even lower angle of $1.15\degree$, 
correlated gaps 
also begin to emerge on the electron-doped side after $\gamma$LL$_0$ is merged with conductance band VHS
(black arrows in \prettyref{fig: fig2}h). %
These correlated gaps correspond to Chern insulators that 
emanate from integer filling factors when electron-electron interactions are strong compared to the width of the Chern bands in finite magnetic fields\cite{saitoHofstadterSubbandFerromagnetism2020,choiCorrelationdrivenTopologicalPhases2021}. %
Our observations suggest that such correlated phases emerge only once portions of the bands around $\gamma$ become maximally flat and join with the VHS. %

To explore the development of zero-field correlated phases 
as a function of
twist angle, we perform angle-dependent 
gate spectroscopy to trace out the evolution of the LDOS at the Fermi level\cite{choiCorrelationdrivenTopologicalPhases2021} ($V_{\rm{Bias}}\approx0$) as a function of charge density ($V_{\rm{Gate}}$); 
 see \prettyref{fig: fig3}a-b. Pronounced LDOS suppression near the Fermi energy occurs at 
certain integer filling factors $\nu$. %
For all angles, prominent suppression is observed at $\nu = \pm4, 0$, reflecting the small LDOS around the 
CNP ($\nu = 0$) and 
band gaps at full fillings ($\nu = \pm4$). At $\nu = \pm2$, $+3$, $+1$ 
we additionally observe sharp LDOS drops that can be attributed to emerging correlated gaps similar 
to those resolved in transport\cite{caoCorrelatedInsulatorBehaviour2018, yankowitzTuningSuperconductivityTwisted2019,luSuperconductorsOrbitalMagnets2019, 
aroraSuperconductivityMetallicTwisted2020}. 
Importantly, the observed LDOS suppressions at integer fillings begin to emerge
within the same range of angles that displayed considerable band flattening in finite 
fields (\prettyref{fig: fig2}, \prettyref{fig: fig3}a and \prettyref{exfig: integer_LDOS_suppression}). 
The angle onsets of the insulating regions for the conduction and valence bands 
(marked by dashed lines and arrows) have an electron-hole asymmetry that is also 
consistent with the band flattening. Spectra at $\nu = \pm2$ taken at various twist angles 
(\prettyref{fig: fig3}c,d) indeed show that LDOS suppressions originate from the 
development of a gap at the Fermi energy, and that the $\nu=-2$ gap emerges at slightly higher 
angles---corroborating electron-hole asymmetry. The maximal size of this half-filling 
gap is $\sim 1.5$~meV, lower than the initial reports 
from spectroscopic 
measurements\cite{choiElectronicCorrelationsTwisted2019,kerelskyMaximizedElectronInteractions2019,
xieSpectroscopicSignaturesManybody2019}, but slightly larger than the activation gap extracted from 
transport\cite{caoCorrelatedInsulatorBehaviour2018, yankowitzTuningSuperconductivityTwisted2019, luSuperconductorsOrbitalMagnets2019, aroraSuperconductivityMetallicTwisted2020}. 
We note that the LDOS suppression from the observed gaps in \prettyref{fig: fig3}c,d may also be, in part, related 
to the Fermi surface reconstruction due to flavour symmetry-breaking cascade\cite{zondinerCascadePhaseTransitions2020, 
wongCascadeTransitionsCorrelated2019} (see also discussion of \prettyref{fig: fig4}). As we show in the following, the band 
flattening also creates conditions that favor the cascade.     

A theory analysis of the continuum-model band 
structure\cite{bistritzerMoireBandsTwisted2011,koshinoMaximallyLocalizedWannier2018} with interactions 
treated at a mean-field level in part accounts for the observed 
band flattening and related symmetry-breaking cascade instabilities near the magic angle 
(see SI, section 6 for details).  
While the doping dependence of the moir\'e band deformation at larger angles is well-modeled by 
assuming only a Hartree correction (\prettyref{fig: fig1}), near the magic angle a more complete 
Hartree-Fock treatment is necessary. For example, 
including only the Hartree term\cite{guineaElectrostaticEffectsBand2018,rademakerChargeSmootheningBand2019,goodwinHartreeTheoryCalculations2020} predicts a dramatic band inversion at the $\gamma$ pocket 
that is not observed experimentally 
(\prettyref{exfig: problems_with_hartree}a-d). The Fock term in part counteracts this band 
inversion---thereby stabilizing the band flattening; see \prettyref{fig: fig3}e-g, \prettyref{exfig: 
problems_with_hartree}e-h and SI, section 3. 
Importantly, our calculations predict 
that near the magic angle, the density of states (DOS) at the Fermi 
level ($\mathrm{E_F}$) is significantly amplified (by up to a factor of $\sim4$ for $\nu=2$ and $\sim15$ for $\nu=3$) relative to expectations from non-interacting models (\prettyref{fig: fig3}h). 
This considerable interaction-driven DOS increase magnifies the 
already high DOS around the VHS and accordingly promotes the symmetry-breaking cascade of electronic 
transitions\cite{zondinerCascadePhaseTransitions2020}. 
For a more quantitative treatment of the cascade instabilities, we compare the energies of unpolarized 
and flavor-polarized states at integer fillings for different twist angles.  
In a non-self-consistent treatment that neglects band renormalization 
(see \prettyref{exfig: problems_with_hartree}), depending on the model, either no 
cascade is expected, or flavor polarization is preferred only in a narrow window around the magic angle.
A self-consistent treatment that incorporates band flattening (\prettyref{fig: fig3}i), 
by contrast, predicts cascading over broader twist-angle windows whose widths depend 
strongly on filling---in agreement with experimental data on the electron-doped side 
of \prettyref{fig: fig3}a (see also \prettyref{exfig: dos_ef_absolute units}). 

Our observations reconcile the apparent discrepancy between the emergence of correlated phases 
around the magic angle and experimental observations at odds with seemingly crucial aspects 
of the band theory predictions: (i) large Dirac velocity around the 
CNP\cite{caoCorrelatedInsulatorBehaviour2018, choiCorrelationdrivenTopologicalPhases2021, parkFlavourHundCoupling2020}, 
(ii) total bandwidth exceeding 40~meV\cite{kerelskyMaximizedElectronInteractions2019, 
choiElectronicCorrelationsTwisted2019,xieSpectroscopicSignaturesManybody2019, jiangChargeOrderBroken2019,zondinerCascadePhaseTransitions2020, 
tomarkenElectronicCompressibilityMagicAngle2019}, and 
(iii) large separation of the VHSs\cite{kerelskyMaximizedElectronInteractions2019, 
choiElectronicCorrelationsTwisted2019,xieSpectroscopicSignaturesManybody2019, jiangChargeOrderBroken2019}. These quantities, Dirac velocity, total bandwidth and the VHS separation, appear not to be essential; for example \prettyref{exfig: vhs_dirac} shows that 
measured Dirac velocity and VHS separation are even smaller at 1.18$\degree$ than 1.04$\degree$, while signatures of strongly correlated 
phases (e.g. symmetry-breaking cascade and correlated gaps) are only present in 1.04$\degree$ comparing these two angles. Instead, we identify 
the interaction-driven flattening of the moir\'e bands around the $\gamma$ pockets, with the consequent increase in the density of states, 
as the decisive feature needed for the formation of correlated phases.

In addition to correlated gaps pinned to integer fillings $\nu=\pm 2, +1,+3$, we also observe 
several other interaction-driven features near the magic angle that have not been discussed 
in previous STM measurements (\prettyref{fig: fig4}). For example, prominent 
LDOS suppressions at the Fermi level are visible in both the $2<\nu<3$ and $-3<\nu<-2$ doping 
regions (\prettyref{fig: fig4}b-d). 
Of the two regions, the feature between $\nu = +2$ and $+3$ has a 'tilted V' shape at small bias voltages, which can be largely understood as a 
consequence of the relative prominence of the flavour polarization in the conduction band compared to the valence band\cite{choiCorrelationdrivenTopologicalPhases2021} (see also 
\prettyref{fig: fig3}a-d). 
There, two of the four flavors are pushed away from the Fermi energy by strong
interactions\cite{zondinerCascadePhaseTransitions2020}, consistent with the LDOS being predominately shifted to higher 
energy (\prettyref{fig: fig4}a,c) and resulting in a highly asymmetric spectrum, as seen in compressibility measurements\cite{zondinerCascadePhaseTransitions2020}. 
In contrast, the spectrum between $\nu = - 2$ and $-3$ exhibits a slightly different shape that 
can not be fully explained by a simple flavour-symmetry-breaking cascade that produces large overall asymmetry around the Fermi 
energy. Additionally, this doping range also 
shows a clear, more symmetric gapped feature at small bias voltages (\prettyref{fig: fig4}e). 
The corresponding gap, spanning almost the entire filling range $-3<\nu<-2$ (\prettyref{fig: fig4}d), reaches its 
maximal size of $2 \Delta \approx 1.1$~meV. Also, it becomes prominent 
only below $\theta=1.16\degree$ (see \prettyref{exfig: integer_LDOS_suppression}), and gradually recedes with temperature and disappears above $5-7$ K (\prettyref{fig: fig4}e). 
The gap size as well the temperature dependence of this feature is similar to the gap at $\nu=-2$; 
however, the filling range observed here is unusually large and cannot be explained simply by 
non-interacting effects (see \prettyref{exfig: trivial_gap}). Furthermore, 
the gap extends over a filling range comparable to that at which superconductivity has been observed in 
transport at similar angles for TBG on WSe$_2$\cite{aroraSuperconductivityMetallicTwisted2020} as well 
as in many hBN-only encapsulated devices\cite{caoUnconventionalSuperconductivityMagicangle2018, 
yankowitzTuningSuperconductivityTwisted2019, luSuperconductorsOrbitalMagnets2019}. This correspondence 
indicates that the observed feature may be related to superconductivity itself or signals the possible 
existence of a pseudo-gap phase that precedes 
superconductivity\cite{hofmannSuperconductivityPseudogapPhase2020}. Regardless of its exact origin, the 
pronounced suppression (instead of increase) of the LDOS near the Fermi energy, together with symmetry-breaking cascade features, for fillings where 
superconductivity is anticipated may suggest either an electronic pairing 
origin\cite{bernevigTBGExactAnalytic2020} or a regime of strong-coupling superconductivity as 
recently pointed out for magic-angle trilayer graphene\cite{parkTunablePhaseBoundaries2020, haoElectricFieldTunable2020}. 
Another interesting observation is that, while the gap-like features at the Fermi energy 
become weaker with increasing temperature, features at higher energies---previously identified to be 
related to symmetry-breaking cascade transitions\cite{wongCascadeTransitionsCorrelated2019, nuckollsStronglyCorrelatedChern2020, choiCorrelationdrivenTopologicalPhases2021}---are enhanced, 
see the temperature evolution in 
\prettyref{fig: fig4}f-i. The relation of these features and various recently reported phases that emerge at elevated temperatures\cite{saitoIsospinPomeranchukEffect2020, rozenEntropicEvidencePomeranchuk2020}
remains a subject for future investigations.

\noindent {\bf References:}

\nocite{*}

\noindent {\bf Acknowledgments:} We acknowledge discussions with Francisco Guinea, Felix von Oppen, and Gil 
Refael. {\bf Funding:} This work has been primarily supported by NSF grants DMR-2005129 and DMR-172336; 
and Army Research Office under Grant Award W911NF17-1-0323. 
Part of the STM characterization has been supported by NSF CAREER program (DMR-1753306). 
Nanofabrication efforts have been in part supported by DOE-QIS program 
(DE-SC0019166). S.N-P. acknowledges support from the Sloan Foundation. 
J.A. and S.N.-P. also acknowledge support of the Institute for Quantum Information and 
Matter, an NSF Physics Frontiers Center with support of the Gordon and Betty
Moore Foundation through Grant GBMF1250; C.L. acknowledges support from the Gordon
and Betty Moore Foundation’s EPiQS Initiative, Grant GBMF8682. A.T. and J.A. are grateful for the 
support of the Walter Burke Institute for Theoretical Physics at Caltech. Y.P. acknowledges support from 
the startup fund from California State University,  Northridge. Y.C. and H.K. acknowledge support from the 
Kwanjeong fellowship. 

\noindent {\bf Author Contribution:}  Y.C. and H.K. fabricated 
samples with the help of R.P. and Y.Z., and 
performed STM measurements. Y.C., H.K., and S.N.-P. analyzed the data. 
C.L. and Y.P. implemented TBG models. C.L., Y.P., and A.T. provided theoretical 
analysis of the model results supervised by J.A. S.N.-P. supervised the project. 
Y.C., H.K., C.L., Y.P., A.T., J.A., and S.N.-P. wrote the manuscript with input from other authors. 

\noindent {\bf Data availability:} The data that support the findings of this 
study are available from the corresponding authors on reasonable request.

\clearpage

\clearpage

\clearpage

\begin{figure}[ht]
\begin{center}
    \includegraphics[width=14cm]{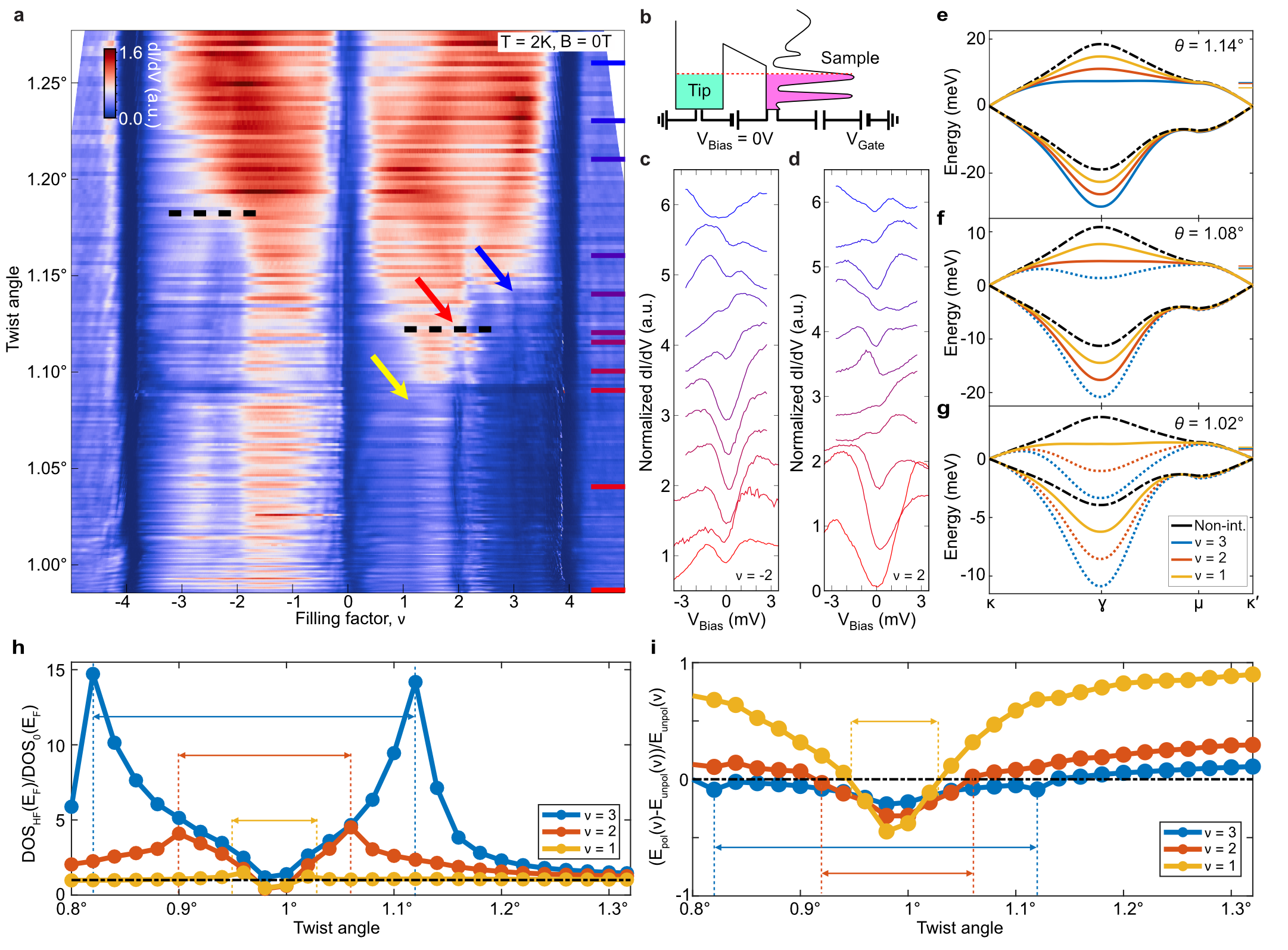}%
\end{center}
\caption{{\bf Emergence of zero-field correlated gaps and symmetry-breaking cascade.}
{\bf a}, Angle and filling-factor dependence of $\mathrm{dI/dV}$ near $\mathrm{E_F}$ 
($\mathrm{V_{Bias}}=0.3$~mV) at $\mathrm{B=0}$~T taken on the same area as in 
\prettyref{fig: fig2}a,b. 
Correlated gaps at $\nu = -2, +1, +2, +3$, observed as an 
abrupt drop in $\mathrm{dI/dV}$ (i.e., LDOS), develop only below certain twist 
angle---in contrast to regions between the flat and remote bands ($\nu=\pm4$) where 
LDOS is small for any angle. Black lines near $\nu = \pm2$ mark the upper bound 
where gaps begins to emerge, while colored arrows indicate corresponding LDOS suppression 
regions at $\nu=+1,+2,+3$. {\bf b}, Schematic showing how data 
in ({\bf a}) is taken: $\rm{V_{Bias}\approx 0}$ is fixed so that the STM tip probes LDOS near the Fermi energy
while $\rm{V_{Gate}}$ is swept. {\bf c, d}, Spectroscopy for $\nu = -2$ ({\bf c}) 
and $\nu = +2$ ({\bf d}) at different local twist angles ranging $\theta = 1.26\degree-0.98\degree$, taken at AB sites. Color coding of the lines correspond to the angles marked by horizontal bars in ({\bf a}). Clear correlated gaps that open only at the Fermi energy are observed only below a certain angle (small wiggles above this angle originate from trivial origins, see \prettyref{exfig: trivial_gap}). Each spectrum is normalized by an average $\mathrm{dI/dV}$ value and offset for clarity. 
{\bf e-g}, Interaction-renormalized band structures at different integer fillings, calculated assuming 
unpolarized ground states (see main text and SI, section 6).  Horizontal lines indicate the relevant chemical potentials.   In cases where polarized states are favorable, dotted 
lines are used.    
The non-interacting band structure is shown in black. {\bf h}, Twist-angle dependence of the DOS at $\mathrm{E_F}$ 
obtained from the interaction-corrected  unpolarized band structure,  normalized by the non-interacting DOS. 
Peaks signal maximal band flattening as seen in {\bf e-g}. 
{\bf i}, Relative energy change for polarized (cascaded) states relative to 
unpolarized states.  Interaction-driven band flattening significantly extends the range of angles, marked by arrows in ({\bf j}) and ({\bf i}), where this relative energy change is negative and polarization becomes energetically favourable (see SI, section 6).
}
\label{fig: fig3}
\end{figure}

\clearpage

\begin{figure}[ht]
\begin{center}
    \includegraphics[width=16cm]{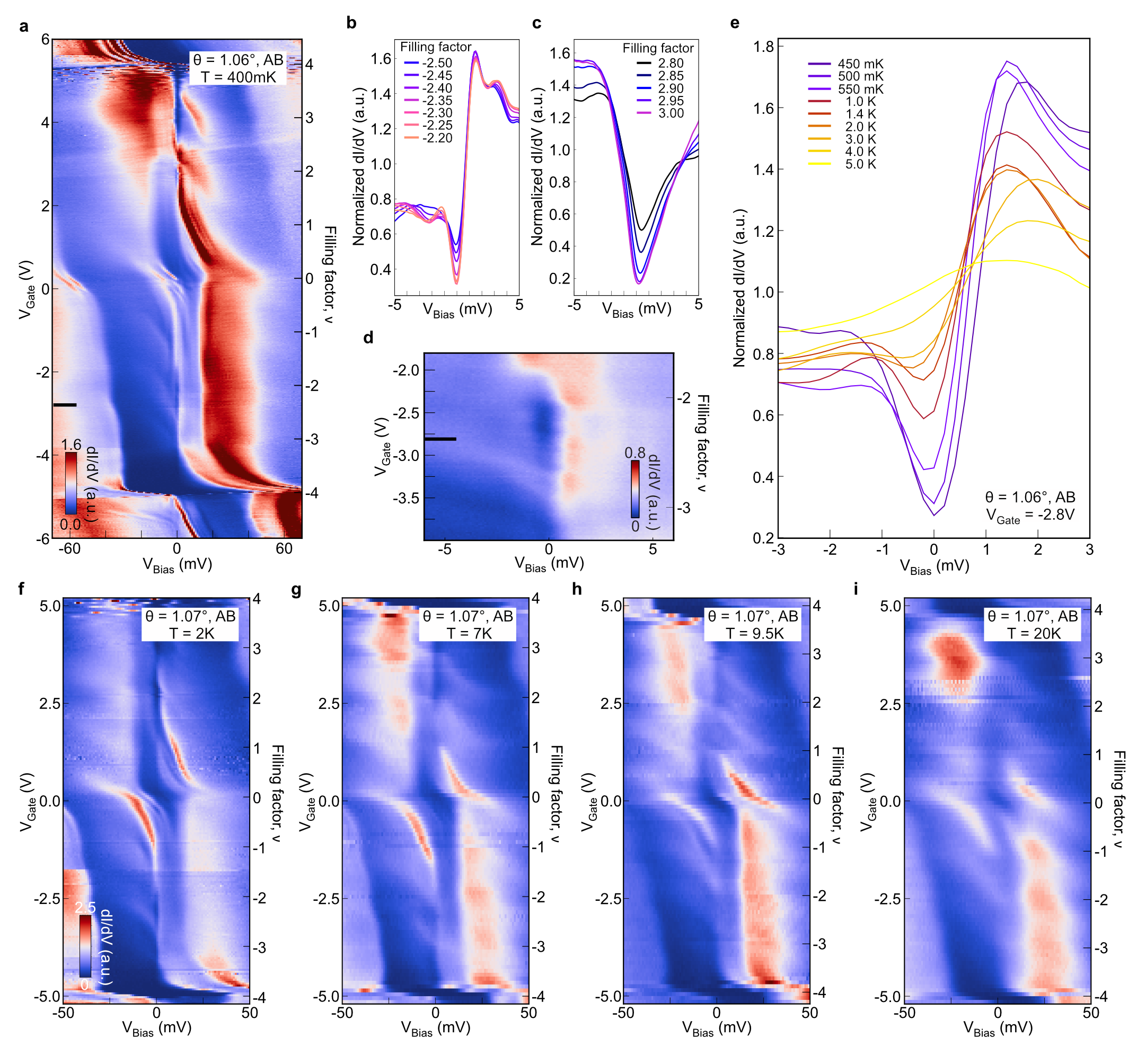}
\end{center}
\caption{{\bf Temperature dependence of correlated gaps around $\mathbf{\nu=\pm2}$ and the symmetry-breaking cascade}. {\bf a}, Point spectra as a function of $\mathrm{V_{Gate}}$ for $\theta=1.06\degree$ 
at $\mathrm{B = 0}$~T. {\bf b,c}, $\mathrm{dI/dV}$ spectra for filling factors ranging from $\nu = -2.5$ to $\nu = -2.2$ 
({\bf b}) and from $\nu = 2.8$ to $\nu = 3.0$ ({\bf c}). {\bf d}, High-resolution point spectra 
of ({\bf a}) focusing on the soft gap between $\nu = -2$ and -3. {\bf e}, $\mathrm{dI/dV}$ spectra at 
$\mathrm{V_{Gate}=-2.8}$~V (as indicated by black lines in ({\bf a,d})) for temperatures ranging from $450$~mK 
to $5$~K at the 
same tip location as ({\bf a}).  {\bf f-i}, Point spectra as a function 
of $\mathrm{V_{Gate}}$ for 
$\theta = 1.07\degree$ at temperatures $\mathrm{T = 2}$~K ({\bf f}), $\mathrm{T = 7}$~K ({\bf g}), $\mathrm{T = 9.5}$~K ({\bf h}), and $\mathrm{T = 20}$~K ({\bf i}). As temperature increases, the cascade features become more pronounced, and their
onset more closely follows integer filling factors, hinting at a characteristic cascade temperature scale of $\mathrm{T\approx 20}$~K as previously noted\cite{zondinerCascadePhaseTransitions2020}.
}
\label{fig: fig4}
\end{figure}
\beginsupplement

\begin{figure}[h]
\captionlistentry{} 
    \label{exfig: gamma_ll_spectrum}
\end{figure}

\begin{figure}[h]
\captionlistentry{} 
    \label{exfig: gamma_spatial}
\end{figure}

\begin{figure}[h]
\captionlistentry{} 
    \label{exfig: displacement_field}
\end{figure}

\begin{figure}[h]
\captionlistentry{} 
\label{exfig: gamma_mapping_conduction}
\end{figure}

\begin{figure}[h]
\captionlistentry{} 
\label{exfig: chern_example}
\end{figure}

\begin{figure}[h]
\captionlistentry{} 
\label{exfig: integer_LDOS_suppression}
\end{figure}

\begin{figure}[h]
\captionlistentry{} 
\label{exfig: vhs_dirac}
\end{figure}

\begin{figure}[h]
\captionlistentry{} 
\label{exfig: trivial_gap}
\end{figure}

\begin{figure}[h]
\captionlistentry{} 
\label{exfig: dos_ef_absolute units}
\end{figure}

\begin{figure}[h]
\captionlistentry{} 
\label{exfig: problems_with_hartree}
\end{figure}

\end{document}